# Wind Energy in the United Kingdom:
## Modelling the Effect of Increases in Installed Capacity on Generation Efficiencies

Anthony D Stephens and David R Walwyn
Correspondence to tonystephensgigg@gmail.com

**Abstract**

The decision by the government in December 2007 that the United Kingdom (UK) should build a 33 gigawatt wind fleet, capable of generating about 10 gigawatts or 25% of the country's electricity total requirement, was a controversial one. Proponents argued that it was the most attractive means of lowering the country's greenhouse gas emissions, whereas opponents noted that it would result in an unnecessary and burdensome additional expense to UK industry and households. Subsequently there have been calls for the wind fleet target to be further increased to perhaps 50% of demand. Although the National Grid has had little difficulty in accommodating the current output of about 10% of the total demand on the grid, this will not be the case for a substantially larger wind fleet. When the wind blows strongly, turbines shed wind/energy which is surplus to demand, leading to significant reductions in generating efficiencies. The purpose of the research described in this paper has been to develop a method for investigating the likely performance of future large UK wind fleets. The method relies on the use of mathematical models based on National Grid records for 2013 to 2015, each year being separately analysed. It was found that the three models derived using 2013, 2014 and 2015 data were sensibly the same, despite a 30% increase in installed capacity over this period. Importantly the predictions were either relatively insensitive to, or could compensate for, perturbations likely to be seen by the grid in future, indicating that the model from a single year's records should have wide applicability as a predictive tool. Accordingly the 2014 data was used to investigate the relationship between wind fleet capacity and energy output, showing that the incremental load factor of the wind fleet will be reduced to 63% of its current level should the wind fleet increase from its current size of 14GW$_c$ (installed capacity) to 35GW$_c$, assuming a base load of 15GW. The model also provides a quantitative relationship between the size of the wind fleet and the reduction in carbon dioxide emissions, which suggests that the maximum contribution from a future UK wind fleet is likely to be a reduction of about 80 million tonnes of carbon dioxide per annum.

# 1. Introduction

On 10th December 2007 John Hutton the UK Business Secretary announced that the government would permit the creation of 33 gigawatt ($GW_c$) off-shore wind capacity[1], to deliver $10GW_e$ or about 25% of the United Kingdom's (UK's) electricity needs (Goodall, 2007). At the time, the decision generated much controversy. Typical of the opposing views were those expressed in a BBC Radio interview by Sir David King, Chief Scientific Advisor to the government from 2000 to 2007, and Maria McCaffery, Chief Executive of the British Wind Energy Association (BBC News, 4 September 2008). The former claimed that:

> "if we overdo wind we are going to put up the price of electricity and that means more people will fall into the fuel poverty trap….. the numbers are difficult to estimate but half a million are not at all unrealistic … as someone who feels we need to reduce our greenhouse gas emissions very substantially in my view it is an expensive and not very clever route to go for that 35 to 40% on wind turbines".

Ms McCaffery took the opposite view:

> "We don't have to pay for wind power; it comes naturally and is totally sustainable ... The expectation is that it will in time drive down the basic cost of energy and actually help the fuel poverty situation, that is certainly our expectation".

A government spokesman said it believed that although the target was ambitious, the government was fully committed to meeting it and the impact on energy bills in the short term would be small (BBC News, 4 September 2008).

In 2009 the European Energy Directive 2009/28/EC set the UK a target of meeting 15% of its energy needs from renewable sources by 2020. To meet this target the UK Renewable Energy Strategy Plan 2009 set a target of 30% of the UK electricity being from renewable sources by 2020, the anticipated wind capacity then being $27GW_c$ ($14GW_c$ on-shore and $13GW_c$ off-shore). The debate has since continued, and in January 2012 a research note by Policy Exchange said that:

> "although the government had claimed that renewable energy policies would actively reduce energy bills by 7% by 2020 compared to what they would have been without policies…. this claim fails the test of clarity… A number of the biggest policies which householders are paying for are hugely and unnecessarily expensive ways of delivering emission reductions… Policy Exchange estimates that the full impact of renewable energy subsidies on average household bills by 2020 (through bills, tax and costs of products and services) to be £400 per year."
> (Less, 2012)

The result was immediately refuted by Chris Huhne, the Minister responsible for the Department of Energy and Climate Change, who stated that:

> "this is nonsense on stilts… overall the impact of our policies on bills by 2020 is estimated to **cut** bills by 7 %".
> (O'Brien, 2012)

Since the original announcement, the size of the UK wind fleet has grown on average by 35% each year, reaching an installed capacity in June 2016 of $14GW_c$, equivalent to about 11% of the total electrical energy demand. From the earliest years of wind power, a key issue has been the eventual size of the fleet, and in 2014 the Royal Academy of Engineering (RAE) summarised the views of a number of consultants regarding the likely size of the wind fleet in 2030. The latters' predictions ranged from 34.4GW$_c$ to 75.3GW$_c$. The RAE commented that:

> "50GW$_c$ would represent levels unprecedented in any system and raise serious issues of managing the system".
> (Royal Academy of Engineering, 2014).

---
[1] In this article, gigawatt is abbreviated to GW, the suffix c is used to denote nameplate or installed capacity and the suffix e denotes delivered power.

This article shows that large wind fleets will inevitably generate short term surpluses which cannot be accommodated by the grid. An important issue is whether such surpluses may be used beneficially or will need to be shed. A review of possible present and medium to long term energy storage and inter-country transfer technologies comes to the conclusion that excess wind generation will be shed, leading to a progressive reduction in efficiency as the wind fleet increases in size.

The main objective of the research described in this article has been to use modelling of real time grid records to quantify the extent of this efficiency loss, expressed relative to the present performance. The challenge generally for this type of analysis is to model fluctuating wind patterns and variable demand. Profiles for supplied wind energy have been constructed from the published data for 2013, 2014 and 2015, resulting in the development of an intuitive model covering the relationships between installed capacity, efficiency, environmental impact and energy security.

## 2. Overview of the United Kingdom's Electricity Generation

The UK energy grid draws on a diverse range of sources, the trends of which for the years 2010 to 2015 are summarised in Figure 1. An important and distinctive trend has been the growing use of renewable sources, with reliance on coal and gas falling from 75% of the total electrical energy supplied in 2010 to 50% in 2015.

**Figure 1. Sources of grid power; 2010 to 2014**

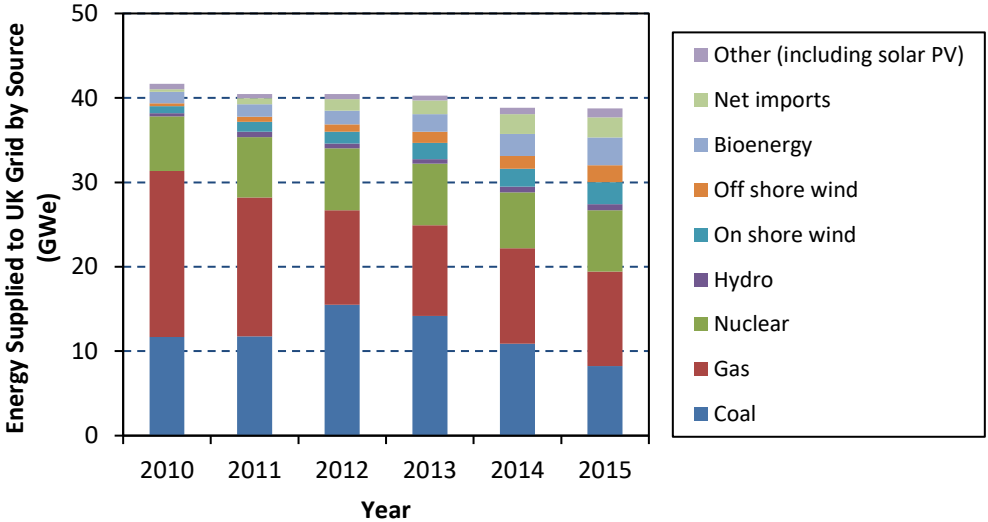

Source: The data was derived from UK government statistics (UK Government, 2016b); the latter are provided in TWh/year and have been converted to the power equivalent via the relationship that 1 TWh/ year is equivalent to 0.11408GWe.

The average weekly demands on the National Grid recorded by Gridwatch during 2013, 2014 and 2015 are shown in Figure 2 (Gridwatch, 2016). Although features such as the cold period in early 2013 and the mild early winter of 2015 are clearly visible, the three curves are generally similar in shape. As we shall see later, this shape similarity allows an important simplification to be made when modelling the generation system.

**Figure 2. Average weekly demand on the National Grid; 2013 to 2015**

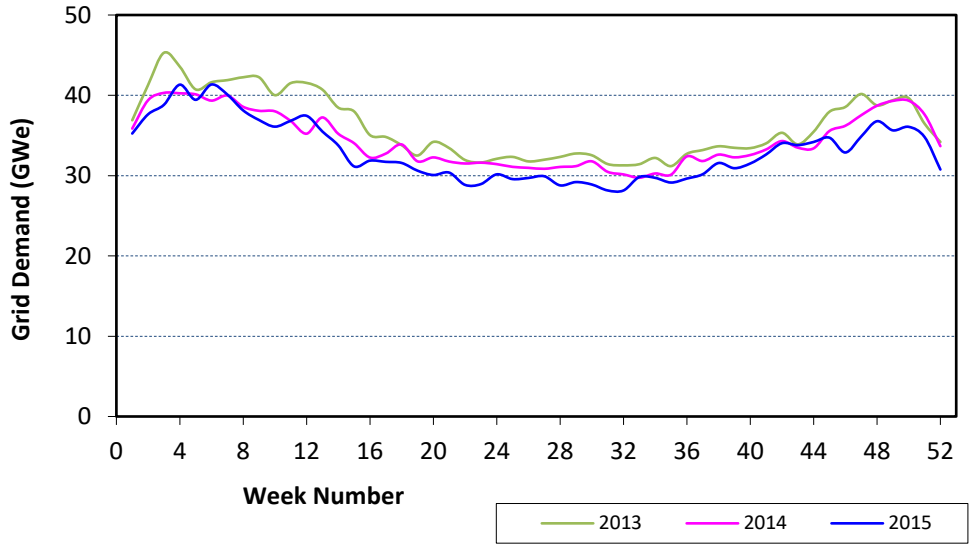

Source: the data have been extracted from Gridwatch (2016) records for 2013, 2014 and 2015. Gridwatch reports in megawatts and are converted here to GW$_e$

Daily demand patterns are shown in Figure 3 and Figure 4. The quasi-sinusoidal demand profile of Figure 3 is typical of summer months; in late autumn, winter, and early spring, on the other hand, the demand pattern shows an additional sharp increase in the evenings (see Figure 4). The weeks illustrated in the figures run from Sunday to Saturday, with reduced demands at the week-ends and also on Monday 26$^{th}$ May in Figure 3, which was a bank holiday.

**Figure 3. Daily demand, coal and gas generation records; week 22, 2014**

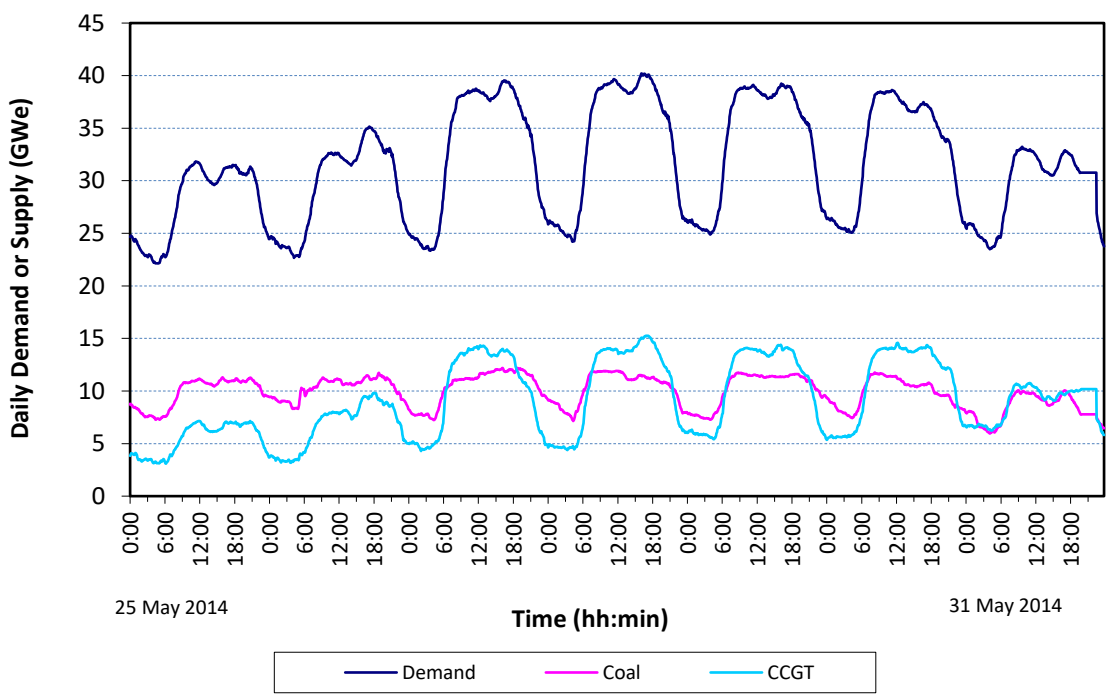

Source: Gridwatch (2016)

**Figure 4. Daily demand, coal and gas generation records; week 45, 2014**

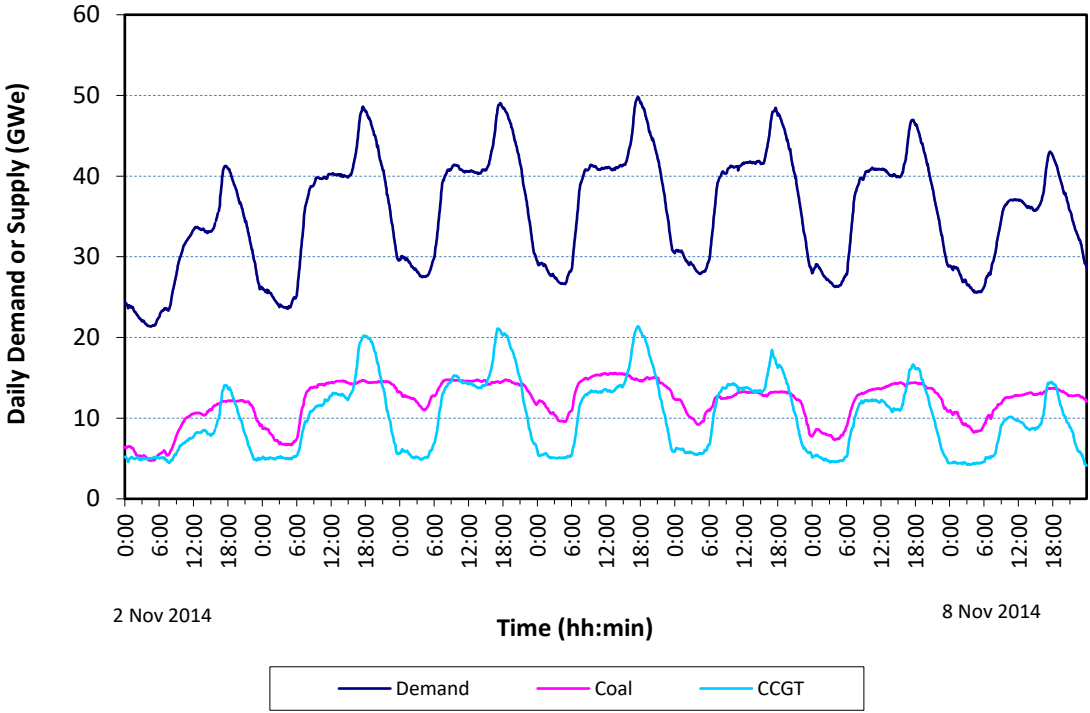

Source: Gridwatch (2016)

Further discussion of the main components of the power generation system now follows.

**2.1   Fossil Fuels (Coal and Gas)**

In 2010, coal and gas provided nearly 75% of the UK's electricity, but as already noted this figure has since fallen to 50% in 2015. Coal and gas are described as 'dispatchable', meaning that they are generation sources which may be increased and decreased on demand. As may be seen in Figures 3 and 4 daily variations in demand has been met largely by adjusting the outputs of gas and coal generation.

Until natural gas became available from the North Sea in 1990, electricity was generated mainly from coal. Gas generation using Combined Cycle Gas Turbines (CCGTs) then progressively replaced coal to become the dominant source of power, but coal generation enjoyed something of a revival as the UK's gas reserves started to become depleted. From a position of being an exporter of gas, the UK became a net importer in 2004, and the first of four Liquid Natural Gas (LNG) import terminals was opened in 2005. An increase in coal generation in 2012 and 2013 at the expense of gas generation which may be seen in Figure 1 was a consequence of Japan buying up a significant proportion of the world's tradeable LNG following the Fukushima nuclear reactor disaster of 2011.

The government announced in November 2015 that coal fired generation would be phased out in the UK by 2025. After that date gas fired generation will become the main source of dispatchable generation. Modern CCGTs may be run up from cold in around half an hour and, although they are inefficient and emit enhanced levels of greenhouse gases on low load, they run close to their maximum efficiency at 60% of full power. They may be ramped up from 60% to full power in around 10 minutes, making them ideal for load following.

## 2.2 Nuclear Power

In recent years nuclear generation has provided just under 20% of grid demand, making it the third largest source of electricity generation after coal and gas. It is unclear however how much longer the UK will be able to rely on nuclear generation. The nuclear fleet currently comprises fourteen Advanced Gas Cooled Reactors (AGR) and one Pressurised Water Reactor (PWR). Although the AGRs are approaching the end of their lives, it is impossible to predict with accuracy when they will be shut down. The life limiting factor is the contraction of the graphite cores caused by oxidation, and only monitoring of the graphite will reveal when the core distortion and hence jamming of the rods in the core channels has reached a level which exceeds an acceptable risk factor. Table 4 in Section 6.3 shows the anticipated AGR shut down dates, as of March 2016. If these estimates are accurate, only the 1.2$GW_e$ Sizewell B PWR reactor of the current reactor fleet will be in service at the end of the 2020s. However the government's aim is to have 16$GW_e$ of new capacity operating by 2030 and the first of some 20$GW_e$ of new generating capacity is expected to be on line by 2025 (World Nuclear Association, 2015). As will be seen later, the amount of nuclear capacity will have a significant impact on the efficiency of the wind fleet as it grows in size.

## 2.3 Biomass

Government subsidies have encouraged a rapid increase in energy generation from biomass in recent years, mainly as a means of meeting the target of generating 25% of UK electricity from renewables by 2020. There has however been a slowdown in the conversion of coal fired to biomass fired generation following questions about whether the 9 to 16 metric tonnes per annum of wood chip, which the UK had intended to source from the US and Canada by 2020, would be able to meet the UK's requirement that the source material should produce less than 200 kg $CO_2$ per GWh (Stephenson and FRS, 2014). Some generation derives from biomass grown in the UK, although it is difficult to understand the ecological logic of burning subsidised maize grown on parts of the UK's prime agricultural land in East Anglia, while at the same time progressively outsourcing agricultural products overseas.

## 2.4 Pumped Storage

Pumped storage plays a small but strategically important role in maintaining the stability of the National Grid. Four storage basins, at Ffestiniog, Cruachan, Foyers and Dinorwic, are filled overnight with up to 30GWh of energy, which is then available for deployment when most needed during the day. The maximum output of the pumped storage system, 2.1$GW_e$, can be deployed in only a few seconds, but pumped storage generation is used sparingly because it is expensive. Not only are capital costs high, but around 25% of the energy needed to raise water into the storage basins is lost on regeneration. Pumped storage energy also provides an important security backup should the grid fail and need access to a reserve of energy.

## 2.5 Imports

Imports are providing the UK with access to overseas generation at a time when its own generating capacity margins are being squeezed because of the shutting down of generating capacity without replacement. Connectors to France and Holland have capacities of respectively 2$GW_c$ and 1$GW_c$, and a new 404 mile interconnector between Revsing in Denmark and Bicker Fen in Lincolnshire with a capacity of 1.4$GW_c$ is due to come online in 2022. Additional links to Norway, Holland and France are also planned (Pagnamenta, 13 June 2016). Interconnectors tend to have capacities of 1 to 2$GW_c$ and, because of their high cost, run at close to maximum capacity transmitting base load generation. As will be discussed later, interconnectors are not suitable for transmitting large short term wind surpluses from one country to another.

In 2015 the French and Dutch interconnectors ran at close to full capacity, helping augment the UK's dwindling dispatchable generating capacity. However there are concerns that this reliance on the European energy system will reduce the long term security of the UK's system, with possible disruptive or complicating factors including (Andrews, 2015):

- the recent shut-down of 8GW$_e$ of German nuclear generation which has changed Germany from being a net exporter of energy to a net importer
- plans to shut-down 2 to 3GW$_e$ of Swedish nuclear capacity
- a growing European reliance on Scandinavian hydro power; although Sweden has 13.5GW$_e$ of hydro generating capacity and Norway 18GW$_e$, and there are plans to expand these capacities further, Scandinavian hydro power cannot necessarily be guaranteed should there be a repeat of the dry winter of 2010/11
- the dependence of France on electricity for heating, rather than gas as in the UK; exceptional winter cold spells could reduce France's ability or willingness to export power to the UK during periods of high demand.

The conclusion would appear to be that interconnectors might be highly beneficial to the UK in normal times, allowing access to low cost generation, but should not be relied upon as back up during times when European generating capacities are at full stretch. The UK's interconnectors with Europe are currently devoted to the transmission of generation from dispatchable sources. In section 3.3 we shall consider whether interconnectors might be appropriate for exporting wind surpluses from possible future large UK wind fleets.

## 2.6   Wind Power

In 2007 the government decided that the UK should build a wind fleet with a capacity of 33GW$_c$ and an estimated output of about 10GW$_e$ (Goodall, 2007). By the end of 2015 the UK wind fleet comprised 6,666 turbines and had a nameplate capacity of 13.61GW$_c$. The National Grid has had little difficulty to date in accommodating the output of the current wind fleet, which has for some years generated around 30% of its nameplate capacity i.e. 0.3GW$_e$ of electrical output per GW$_c$ of capacity, as shown in Figure 5, over a range of wind speeds.

**Figure 5. Wind speeds, fleet efficiencies and capacities; 2001 to 2015**

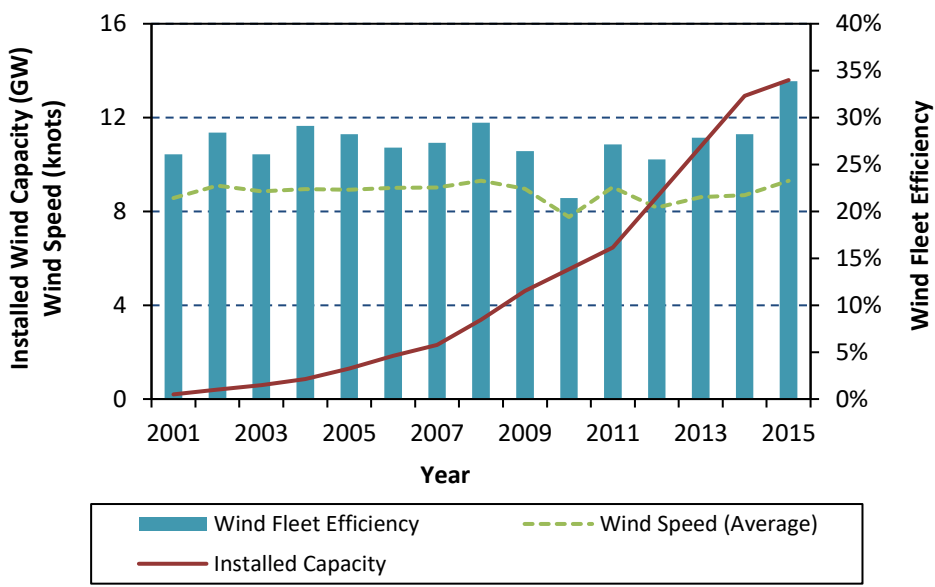

According to the UK Wind Energy Database, if all the turbines under construction or given consent at the beginning of 2016 were built, the result will be a wind fleet of 35GW$_c$. Consultants working for the Department of Energy and Climate Change, whose input has been summarised by the Royal Academy of Engineering, have suggested the possibility of wind fleets ranging in size from 34.4GW$_c$ to 75.3GW$_c$, with a median expectation of 50GW$_c$, by 2030 (Royal Academy of Engineering, 2014). Unfortunately none of the studies appears to have taken into account the implications of wind shedding on the efficiency of large wind fleets as the wind fleet increases in size. In this study, a new approach has been followed to quantify such an effect and we shall present in the analysis that follows the consequences of having wind fleets of up to 80GW$_c$.

The efficiency of UK wind turbines largely depends on whether they are sited on-shore or off-shore. On-shore turbines tend to have average efficiencies measured over a year of about 25%, while off-shore turbines have annual efficiencies of typically 35% (see Figure 6). Because of the large difference between on-shore and off-shore efficiencies the UK government monitors on-shore and off-shore efficiencies as two distinct groups on a quarterly and annual basis. Our model will also assume the wind fleet comprises on-shore and off-shore components whose efficiencies vary from year to year.

There are two complications which need to be addressed when predicting future efficiencies. Firstly, not all of the wind generation is recorded by the grid and monitored by the quasi-real time Gridwatch data base we use for the model. Secondly, in the predictions of the performance of future larger wind fleets we need to take account of the likelihood that a higher proportion of future capacity will be off-shore. Thus, while the proportion of on-shore capacity was 66% in 2014, it is anticipated that this will have fallen to 45% on-shore should the wind fleet reach a capacity of 35GW$_c$. The approach to both complicating factors is described in Section 4.

**Figure 6. Annual load factors of the UK wind fleet 2008 to 2015**

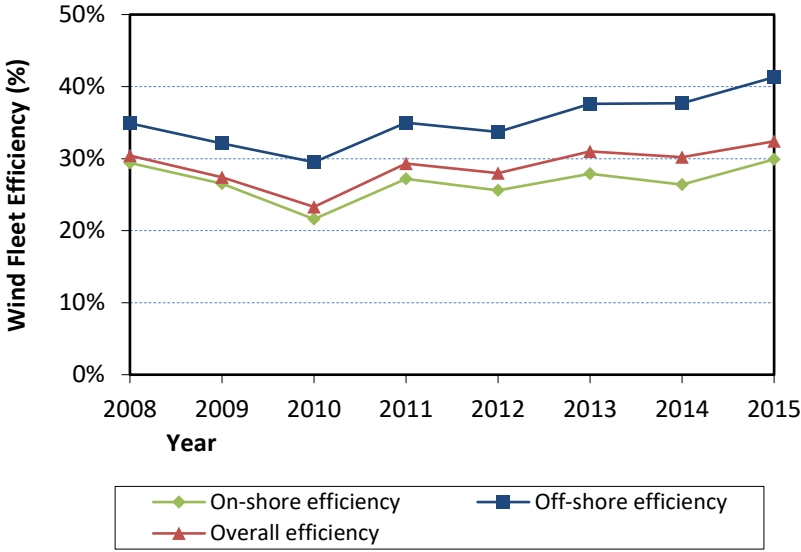

Source: The data was derived from UK Government statistics (UK Government, 2016a)

3. **Managing Wind Surpluses**

Before considering how to model the UK wind fleet, we need first to address how a future grid could cope with wind surpluses. In order to do so, we must first estimate the magnitude of the surpluses likely to arise as a consequence of larger wind fleets, then assess whether current or future technology may be able to handle these surpluses.

Grid records such as those given in Figure 3 and Figure 4 show that daily demand varies by about 7.5GW$_e$ (a 20% variation), equivalent to a total energy requirement of 90GW$_e$h. If a way could be found to store and restore this energy, the energy profile could be levelled and peak demand reduced by 7.5GW$_e$.

Storing the intermittent large surpluses of energy from a large wind fleet would be much more challenging. The simulation results we shall discuss later reveal the occasional surplus of 30 to 40GW$_e$ from a wind fleet of 80GW$_c$, typically occurring once or twice a fortnight and lasting about 24 hours. To store and restore such a surplus would require energy storage of the order of 840GW$_e$h, since these periods of high output can last for 24 hours. Even greater storage would be required to service lengthy wind lulls such as one encountered in weeks 36, 37 and 38 of 2014 (see Figure 10). The model discussed later calculates a deficit for this three week period of 5,374GW$_e$h, even had there been a wind fleet as large as 80GW$_c$.

There are many different means of storing energy, and when we consider their suitability for compensating the intermittent nature of wind generation it is useful to bear in mind the three different storage requirements discussed above, namely 90GW$_e$h to level the daily demand profile and reduce peak demand by about 20%; 840GW$_e$h to store the occasional short term surpluses of 30GW$_e$ to 40GW$_e$ from a 80GW$_c$ wind fleet; and 5,374GW$_e$h to mitigate lengthy wind lulls such as that experienced during weeks 36, 37 and 38 of 2014. It is also important to consider the requirement to transmit the occasional short term surplus generation of 30GW$_e$ to 40GW$_e$.

### 3.1 Technical Solution One; Pumped Storage

The UK's current pumped storage system has an energy capacity of around 30GW$_e$h and a maximum output of 2.1GW$_e$ (MacKay, 2009). It is used daily for a small amount of grid smoothing and in helping the dispatchable generating sources meet peak demands. MacKay (2009) estimated that if all the suitable Scottish lochs were pressed into service as pumped storage reservoirs, they might be capable of storing around 400GW$_e$h of energy. As discussed earlier, pumped storage electricity is expensive, but the potential for 90GW$_e$h of pumped storage capacity to reduce peak demand by 20% and add additional stability in the face of short term wind lulls makes an investment in more pumped storage capacity worth considering. However it is most unlikely that pumped storage will ever be used for the storage of large intermittent wind surpluses, or to mitigate lengthy wind lulls, given the cost.

### 3.2 Technical Solution Two; Battery Storage

In 2009 it was suggested that the conversion of the UK's 20 million cars to electric vehicles (EVs) would allow the storage of 1200GW$_e$h of energy (MacKay, 2009), and in 2016 the Professional Engineer reported a collaboration between Nissan and ENEL of Italy to enable Nissan's EVs to act as electric hubs. These hubs would allow two way flow of energy between the grid and the EVs. An interesting aspect of Nissan's development work is the recycling of lithium-ion electric-car batteries as a means of giving the batteries as 'second-life' as energy storage solutions (Nissan Motor Corporation, 2009). Depending on how they are used, lithium-ion batteries appear to have in-car first lives of around 8 years, after which the battery will have lost 25 to 30% of its original capacity and will be in need of replacement. Instead of discarding these batteries, Nissan aims to use them for a further 5 to 10 years as second-life storage devices. To date, Nissan has made 200,000 electric vehicles, so its second-life batteries should have the potential to store around 4.8GWh of electrical energy.

The United States (USA) would appear to be in the lead in producing lithium-ion batteries for power applications (Atacama, 2016). Tesla Motors, one of the leading USA manufacturers of EVs, which has to date sold around 107,000 EVs in 42 countries is reported to be planning to open a factory in Nevada in 2016 in order to manufacture 500,000 electric vehicles a year within 5 years (Hipwell, 2016). AES Energy Storage (AES), a company active in power applications, is building a 100MW storage device for power smoothing, and has agreed to buy up LG's lithium-ion batteries over a number of years to enable it to build a 1GW$_e$ output storage device for power applications. AES, which has built a 10MW array in Northern Ireland, is one of the bidders for a National Grid contract to install a 200MW battery-powered back-up array (Pagnamenta, 14 July 2016). The National Grid is expected to invite bids later this year for an additional 500MW of energy storage.

It is tempting to regard lithium-ion energy storage as a means of simultaneously solving the problems of $CO_2$ emissions from vehicles, smoothing grid demand and ameliorating wind generation variability. However it is clear that lithium-ion battery manufacture is currently a small scale operation and in its infancy worldwide; its main market is the production of highly priced luxury vehicles for the few who can afford to buy them. Bearing in mind that the leading US manufacturer of electric vehicles only aims to increase its manufacturing capacity to 500,000 units by 2021, and the energy storage of $1GW_e$ is still a distant aspiration, it seems unlikely that lithium-ion technology will have any impact on UK electricity generation in the 2020s. A feasible aim might be to have a sufficient number of electric vehicles connected to the UK grid in the 2030s to achieve a reasonable degree of demand smoothing (around 1.5 million grid connected vehicles would be needed to smooth the demand patterns shown in Figures 3 and 4, thereby reducing peak demand by around $7.5GW_e$). It might be thought that smoothing grid demand would significantly improve the efficiency of large wind fleets but, as we shall see later, the efficiency of a wind fleet is relatively insensitive to the size of the cyclic component of demand.

### 3.3   Technical Solution Three; Inter-Country Connectors

In Section 2.5, the growing role of intercountry connectors as a means of transmitting dispatchable energy between the UK and Europe was discussed. The question of whether interconnectors could also play a useful role in future in transmitting wind surpluses requires consideration of two different issues. The first issue is whether connection to countries with different weather systems gives better security against wind lulls, and the second is whether the surpluses which will inevitably arise from a future large UK wind fleet can be beneficially exported for use elsewhere.

The first question may be addressed by studying the wind records of the UK and European neighbours it might rely on in times of deficit. The report of the Royal Academy of Engineering (2014) on wind energy concluded that the UK's weather is highly correlated with that of Denmark and reasonably well correlated with that of Germany. We would expect therefore to be able to identify records of low wind generation across Europe and this is the case. If the UK had invested in interconnectors with Europe to export/ import wind generation it would almost certainly have wished to use the interconnectors on 13th December 2015, when the UK wind fleet produced only 231GW, its lowest output of the year. However, the weather map for that day showed practically no wind above 4 knots across Europe, Scandinavia, Western Russia, the Middle East and Northern Africa (Wind Finder, 2015). There have been other occasions when record winter demands on the grid have coincided with low generation from both the UK and European wind fleets, including 7th Dec 2010, when the UK's wind fleet load factor was 5.8%, that of Denmark 4% and Germany 3%. As the National Grid commented at the time, the "winter peak normally occurs when temperatures are low and this often results from anti-cyclonic conditions that also mean very little wind .... over a very large area" (National Grid UK, 2009). We may conclude that interconnectors cannot be relied on to compensate for lack of UK wind generation during severe winter wind lulls.

The question of whether interconnectors to Europe would enable UK wind surpluses to be beneficially used elsewhere is rather more complicated; it requires an understanding of both the size of the surpluses likely to arise and whether potential customers are likely to be able to make beneficial use of UK surpluses when they arise. Fortunately the model whose development we shall discuss in the next section helps us address both these questions.

Figure 8 shows predictions of wind generation for week 45 of 2014 had the UK wind fleet been a range of sizes up to $80GW_c$. It may be seen that although a $80GW_c$ wind fleet would have produced a surplus of around $40GW_e$ on 6th November 2014, there would have been deficits during periods of peak demand on 3rd, 4th, and 5th November. The records for the other weeks of 2013, 2014 and 2015 show fairly frequent short term surpluses of $30GW_e$ to $40GW_e$ from a $80GW_c$ wind fleet. Although Figure 8 provides a useful visual snapshot of wind surpluses/deficits for a single week, an economic assessment needs to consider surpluses throughout the year. Figure 12 enables this to be done;

assuming a base generation of 15GW$_e$, a median expectation for 2030, the figure suggests that a wind fleet of 80GW$_c$ would produce an average generation of 26.0GW$_e$ over a year. Of this, 15.2GW$_e$ would have been accommodated by the grid, and 10.8GW$_e$ would have been surplus to the grid's needs. Interconnectors with capacity of 40GW$_e$ would have enabled the surplus to be transmitted for use elsewhere, but the maximum utilisation of the 40GW$_e$ interconnector would have been only 27%. This utilisation estimate assumes that customers would have been able to accommodate all of the UK surplus generation but, because of correlation of wind patterns between the UK and potential customers, interconnector utilisation is likely to be considerably less than 27%. Additionally, we need to consider whether European neighbours would be interested in accommodating highly unpredictable high priced UK wind surpluses. Denmark and Germany already generate a higher percentage of their own electrical demands from wind than the UK, and will already be experiencing problem of accommodating the variability of their own wind. They are more likely to be interested in importing generation which can be relied upon, such as nuclear generation from France, and hydro power from Scandinavia, than additional unpredictable wind power from the UK. A final consideration is that interconnectors are expensive. Even if customers could be found for all of the UK's wind surplus of 10.8GW$_e$, it is difficult to see an economic case being made for investing in 40GW$_e$ of interconnectors which have a maximum utilisation of 27%.

Some mention is needed of Denmark, often referred to as a country whose wind generation credentials should be emulated by others. Denmark produces a higher percentage of its electricity from wind than any other country (42% in 2015), and has 6.5GW$_e$ of DC and AC inter-connectors with its neighbours despite having a home consumption of only 3.9GW$_e$ (Wikipedia, 2016). Some of the inter-connectors have very low utilisations. This highly unusual arrangement is partly because of Denmark's geography, being particularly windy, and having Nordic neighbours who have high rainfall and natural storage basins; exchanging wind and hydro energy makes economic sense for all the parties. Also Denmark, unlike the UK, has for several decades adopted the strategic objective of being a world leader in wind turbine manufacturing. A downside is that Danish households pay considerably more than other European countries for their electricity; about twice as much as UK households. The Danish model only works because it is a small country adjacent to neighbours with larger electricity markets. The model does not work in reverse for large countries with smaller neighbours.

In conclusion, it is most unlikely for the foreseeable future to be economically beneficial for the UK to store wind surpluses or transmit them to other countries. The model which will be discussed in the next section will assume that wind surpluses are shed.

## 4.  Methodology; Modelling the Wind Fleet

### 4.1  Assumptions

As the wind fleet increases in size, it will progressively generate more electricity than is needed by the grid; in the cases we shall consider, the excess generation will frequently be greater than 30GW$_e$ to 40GW$_e$. The conclusion of the discussion in the previous section is that it is unlikely to be economic to store or transmit such surpluses. In what follows it will be assumed that any wind generation surplus to demand, is shed.

In some countries, particularly those with large land masses and varying climactic conditions, wind fleet efficiencies can vary significantly depending on location. However, this is not the case for the UK. As discussed in Section 2.6, government wind statistics assume the UK wind fleet to be two components only, on-shore and off-shore components. As may be seen in Figure 6, efficiencies of on-shore and off-shore wind fleets, and the overall wind fleet efficiency, move largely in synchronism with one another. For modelling purposes we have assumed that the wind fleet is homogeneous with a single average efficiency.

## 4.2 Model Structure and Variables

It was found possible to use a model with only three variables to address the questions of interest to this study, considerably simplifying what might otherwise have been a problem too complex to analyse. The variables chosen were demand on the grid (demand), wind generation, and a variable which is the aggregate of other generation sources that do not need to be considered individually (see Figure 7). This aggregate variable, called "base generation", includes nuclear, biomass, hydro, energy imports and solar generation. It was thought that a single year's Gridwatch data should be sufficient to derive a model of general applicability but, to prove this to be the case, models were developed separately using 2013, 2014 and 2015 data. In this section a model derived using 2014 Gridwatch data is described.

**Figure 7. Dispatchable, wind and base generation; 5 to 7 November 2014**

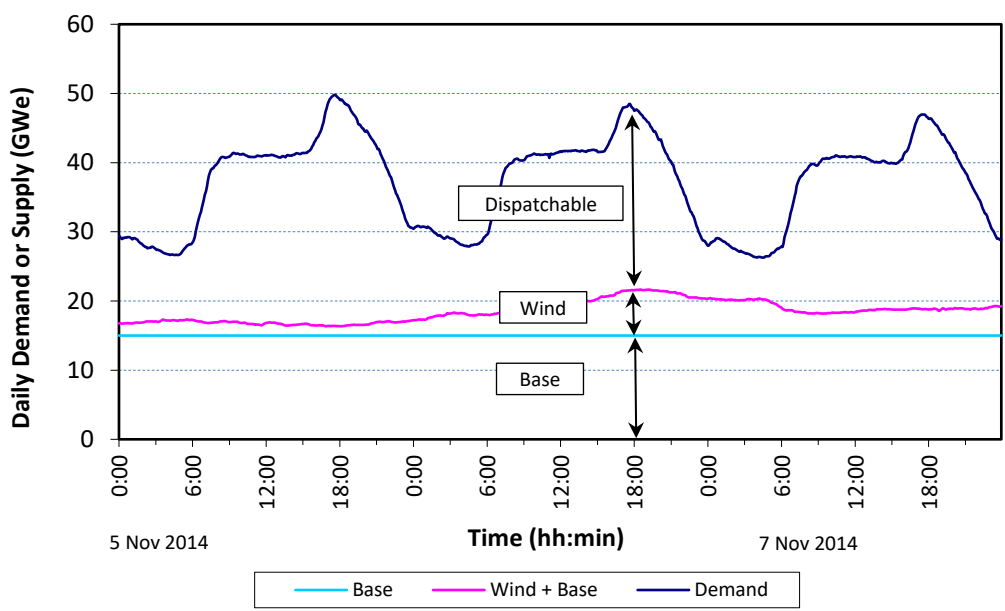

A problem which had to be overcome is that although Gridwatch records all off-shore generation, it records only a portion of on-shore generation. Table 1 shows how this problem was resolved using a combination of UK government and Gridwatch records. UK Government records of on-shore and off-shore capacities at the end of 2013 and 2014 are shown in column 1 and 2. Ideally we would have liked to use week-by-week wind fleet capacities in building our model. Since this information is not available, it was necessary to use annual average wind fleet capacities, the averages capacities for 2014 being shown in column 3. The ultimate justification for making this simplifying assumption is that, as we shall see later, the models derived separately using 2013, 2014 and 2015 data prove to be almost identical.

**Table 1. Estimate of wind fleet capacity as recorded by Gridwatch; 2014**

|  | 2013 ($GW_c$) | 2014 ($GW_c$) | Average ($GW_c$) | Govt. Records ($GW_e$) | Gridwatch ($GW_e$) | Recorded Capacity ($GW_c$) |
|---|---|---|---|---|---|---|
| On-shore | 7.519 | 8.486 | 8.003 | 2.124 | 0.895 | 3.375 |
| Off-shore | 3.696 | 4.501 | 4.099 | 1.530 | 1.530 | 4.099 |
| Total | 11.215 | 12.987 | 12.101 | 3.654 | 2.425 | 7.473 |

Source: UK Government (2016a) and Gridwatch (2016).

The government records of both on-shore and off-shore generation are shown in column 4, and the total generation "seen" by Gridwatch (2.425 GW$_e$) is shown in column 5. Because we know that all of the off-shore generation was "seen" by Gridwatch (1.530 GW$_e$), on-shore generation can be calculated by difference (0.895 GW$_e$). The average annual load factors recorded by the UK Government were then used to estimate the on-shore and off-shore capacities which gave rise to the generation "seen" by Gridwatch, shown in column 6. To predict what the wind generation would have been had the wind capacity been say 10 GW$_c$ rather than 7.473 GW$_c$, each wind record of 2014 was multiplied by a ratio of 10 to 7.473 (1.338). Higher multiples of 1.338 were used to calculate the wind generated by wind fleets ranging in size from 20GW$_c$ to 80GW$_c$.

### 4.3 Extrapolating and Estimating Efficiency

Gridwatch provides wind generation records taken every 5 minutes. The total amount of data downloaded for the three years 2013, 2014 and 2015 for this project was of the order of 1.25 million items. To ease the problems of handling and visualising such large amounts of data, the records of interest (demand and wind) were downloaded a week at a time only, together with the time stamp for the readings. The three columns were copied into an Excel spreadsheet, pre-formatted to carry out the required calculations, so that a year's data generated 52 spreadsheets, each with the same basic format, but with different input data.

For each data set the spreadsheet first calculated what the wind generation would have been had the wind fleet capacity been a range of capacities from 10GW$_c$ to 80GW$_c$ rather than the capacities in existence in 2013, 2014 or 2015. The result is a set of 52x3 graphs such as that illustrated in Figure 8. The model then checked for generations which exceed demand and, when excess generation is found, constrained the output to be equal to demand, as illustrated in Figure 9. The weekly averages were then calculated from the data underpinning Figure 8, and then imported into another spreadsheet to allow a week-by-week display of a year's results. The data for 2014 from the latter spreadsheet are shown in Figure 10.

**Figure 8. Predicted unconstrained wind generation from wind fleets varying in size between 20GWc and 80GWc**

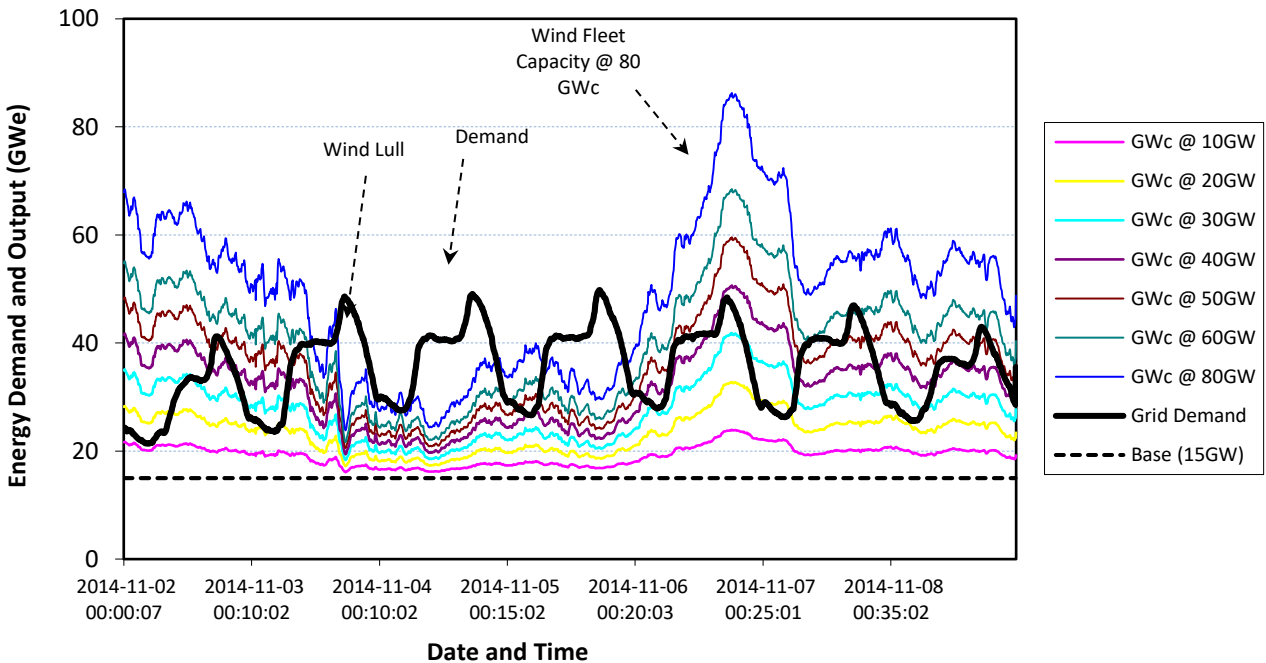

It may be seen from Figure 8 that during week 22 of 2014 wind is just starting to shed for a wind fleet capacity of 20GW$_c$. Had the wind fleet been 80GW$_c$ in size however, although some dispatchable generation would have been required on 3$^{rd}$, 4$^{th}$ or 5$^{th}$ November, wind generation would have exceeded demand by some 40GW$_e$ on the night of 6$^{th}$ November. This change from deficit to surplus over a period of only a few hours illustrates why it is necessary when modelling wind fleet efficiency to use real time data. The spreadsheet underpinning Figure 8 calculates, for each 5-minute time interval, how much wind generation could have been usefully used. Wind generation is constrained to lie on or below the demand line as shown in Figure 9.

**Figure 9. Predicted useful wind generation from wind fleets varying in size between 20GWc and 80GWc**

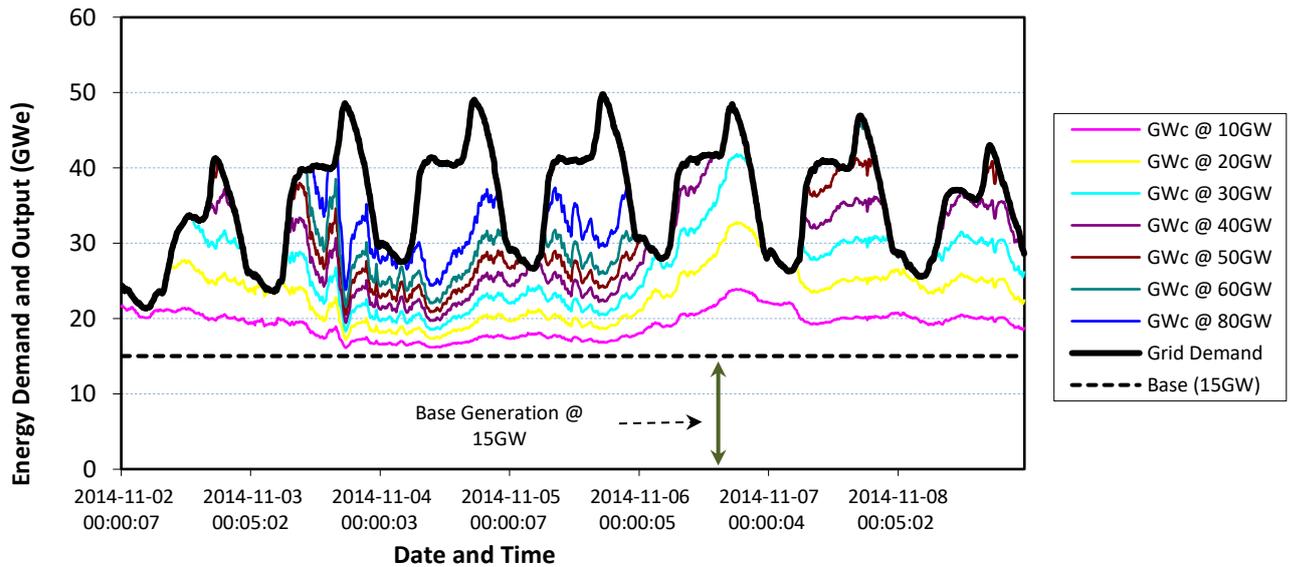

**Figure 10. Predicted average weekly wind generations for different wind fleet capacities; 2014**

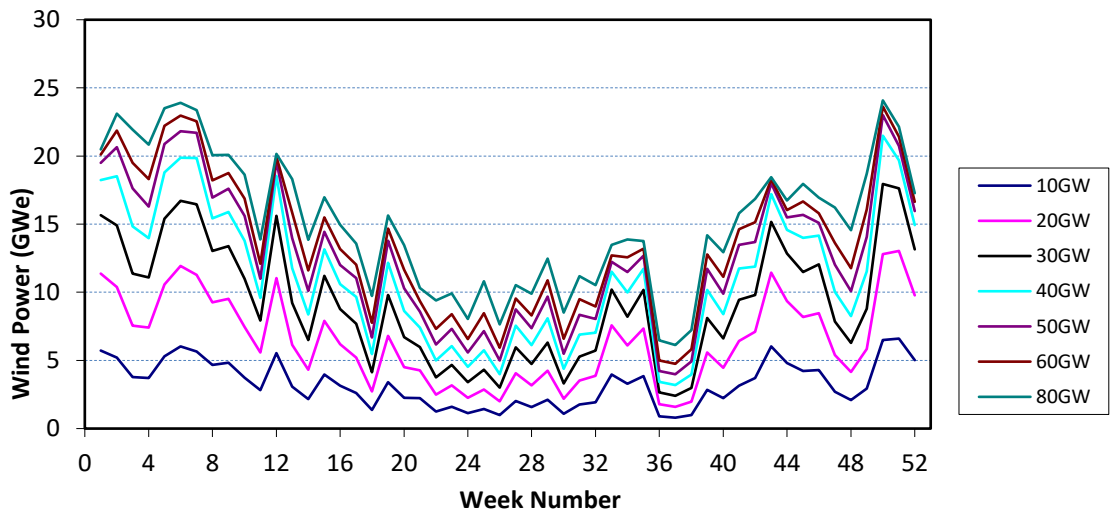

The yearly averages for each wind fleet capacity were then calculated from the data underpinning each week and used to derive the wind fleet efficiency curves for each year, as shown in Figure 11. It is noted that the closeness of the three curves to one another justifies the original supposition that it might be possible to develop a model of general applicability from a single year's grid records only.

**Figure 11. Wind generation curves, GWe vs. GWc, calculated; 2013 to 2015**

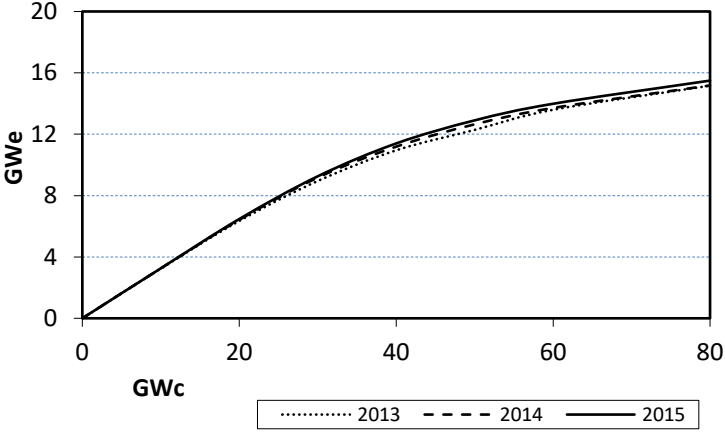

When simulating complex systems, it is useful to be able to check results by independent means. Processing the grid records a week at a time, not only eases the problem of data handling, but also allows visualisation of the intermediate stages of calculation. It increases the confidence in the results by identifying, for example, the abnormally cold and mild winters as shown in Figure 1.

**4.4    Compensating for Different Average Annual Demands**

The average demands for 2013, 2014 and 2015 derived from the Gridwatch data may be deduced from Figure 1, and are summarised in Table 2. Since Figure 1 shows that the pattern of weekly variations is similar for the three years, albeit at slightly different levels, a simple method of compensating for the different average demands is to adjust the base generation by the same amount. Table 2 shows how a base level of 15GW$_e$ for 2014 was adjusted for 2013 and 2015 to take account of the different levels of average demand in 2013 and 2015. These adjusted base levels were used by the models to calculate the annual efficiencies.

**Table 2. Compensating for differences in average demand by adjusting the level of base generation**

|  | 2013 | 2014 | 2015 |
|---|---|---|---|
| Average demand (GW$_e$) | 36.075 | 34.409 | 33.007 |
| Difference from 2014 (GW$_e$) | +1.666 | 0 | -1.334 |
| Adjusted base level (GW$_e$) | 16.666 | 15.000 | 13.666 |

**5.    Extending the Applicability of the Wind Generation Curve**

The generation curves of Figure 11 are sensibly the same, regardless of whether 2013, 2014 or 2015 grid records are used in their generation. In what follows we shall use the GW$_e$ vs. GW$_c$ curve generated using 2014 records. We now need to discuss the circumstances which might require modifications to be made to the GW$_e$ vs. GW$_c$ curves, namely different levels of base generation; different patterns of daily demand; different ratios of on-shore to off-shore capacity; and different load factors.

**5.1    Different Levels of Base Generation**

Base generation was roughly 13GW in 2014, but it is impossible to predict with any accuracy what the base generation will be in future. Some of the components of base generation such as bio energy have been increasing rapidly in recent years, but nuclear generation might be considerably lower or higher in 2030 than its level of 6.61GW$_e$ in 2014. In view of the high degree of uncertainty about the future level of base generation it is necessary to recalculate the GW$_e$ vs. GW$_c$ curves for a range of base generation levels. The results are displayed in Figure 12 in the same format as Figure 11.

Compensation for different levels of average annual demand might be complicated if it were not for the fact that the patterns of demand shown in Figure 2 are largely seasonal, with slight differences in annual averages. Since the level of annual demand and base generation restrict the area in which the wind fleet may operate to a similar way, it is justifiable to compensate for small changes in average demand by adjusting base generation with the same amount.

## 5.2 Different Patterns of Daily Demand

The patterns of daily demand in 2014 illustrated in Figures 3 and 4 roughly approximate to average demand which changed from day to day and from week to week, onto which is superimposed a roughly cyclic daily demand of approximately 15GW$_e$ peak-to-peak. It is anticipated that this cyclic component of demand will be reduce in future as smart equipment is progressively attached to the grid in order to reduce peak demand on the grid. Although it would be possible to predict the effect of reducing the cyclic component of demand by inputting different demand patterns into the model, individual simulations would be difficult and would not provide useful general insights.

**Figure 12. The effect of the level of base generation on wind generation for different sizes of wind fleets**

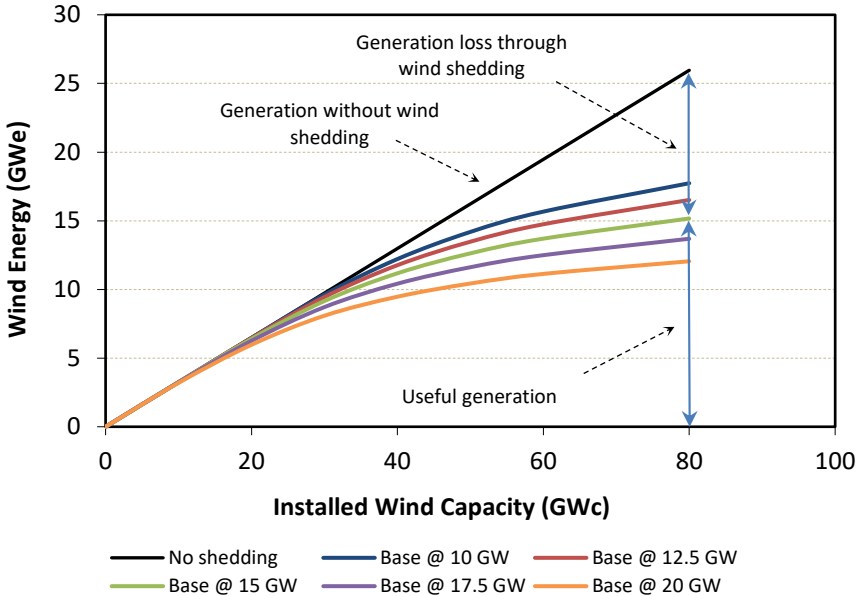

What does provide an extremely useful general insight is to carry out a simulation in which the cyclic component is reduced to zero. This may be carried out easily by replacing the pattern of daily demand seen in 2014 with a pattern of demand held constant each week, at the level shown in Figure 1, which is close to the pattern we would expect if, at some time in the distant future, sufficient smart equipment was attached to the grid to create a level demand throughout each day. The result in shown in Figure 13, which compares the wind generation predictions using 2014 grid data (without markers) and with demand modified to eliminate the cyclic component (with markers).

The initially surprising feature of Figure 13 is that although different levels of base generation have a first order effect on the wind generation predictions, large changes in the magnitude of the cyclic component of demand have only a second order effect. The reason for this state of affairs is that reducing the peak-to-peak variation reduces the highest level of demand (during the day) but increases the lowest level of demand (at night). The former causes slightly more wind to be shed during the day and the latter slightly less at night, the two effects largely cancelling each other out. It will be many years, if ever, before sufficient smart equipment is attached to the UK grid to create a demand which has no daily cyclic component. For the foreseeable future it is reasonable for planning purposes to ignore the secondary effect on the GW$_e$ vs. GW$_c$ curve caused by different level in the cyclic component of daily demand.

**Figure 13. Comparison of wind generation predictions using 2014 grid data (lines only) and with no cyclic component (markers only)**

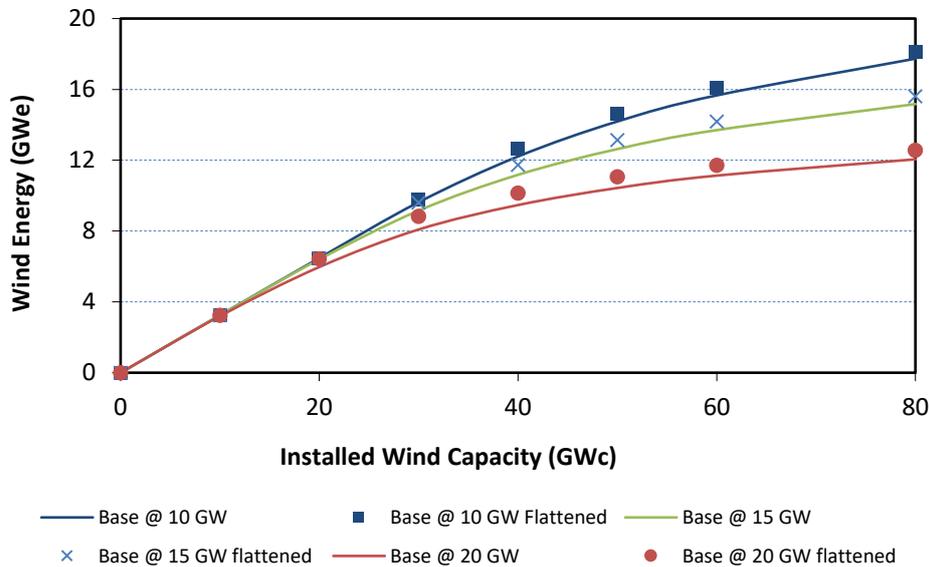

## 5.3 Different Ratios of On-Shore to Off-Shore Capacity and Load Factors

It is a much simpler matter to adjust for different ratios of on-shore to off-shore capacity than different levels of base generation. While compensation for different base generation levels require the rework of the weekly and annual averages to produce new GW$_e$ vs. GW$_c$ curves, compensation for different proportions of on-shore generation merely require adjusting the GW$_e$ value for each GW$_c$ value using the formula:

$$GW_{e,1} = GW_{e,2} \left( \frac{n_2 - p_1(n_2 - n_1)}{n_2 - p_2(n_2 - n_1)} \right)$$

where $p_1$ and $p_2$ are the different proportions of on-shore capacity, and $n_1$ and $n_2$ are respectively the on-shore an off-shore efficiency factors.

In 2014, the proportion of capacity which was on-shore was 0.66, but this is expected to fall to 0.45 by the time the wind fleet increases to 35GW$_c$ (as progressively more off-shore capacity is added). We would therefore expect a requirement to correct GW$_e$ for the reduction in p value by multiplying the values calculated in 2014 by about 7.8%. However although the p value was 0.66 in 2014, the p value of the generation seen by the grid was 0.45. It is not necessary therefore to apply any compensation for the expected generation when the fleet reaches 35GW$_c$ in size.

Although Figure 6 shows the average load factor changed little during 2011 to 2015 we cannot preclude years of abnormal wind conditions, as occurred in 2010. Nor can we preclude the possibilities of technical innovation leading higher load factors or ageing of the wind fleet lower the load factors. Compensation for different load factors merely requires the wind fleet characteristics of Figure 12 to be adjusted in proportion to the different load factors.

## 6. Discussion

### 6.1 Predicting Wind Fleets Efficiencies

Although the critical issue in any study of the optimal wind fleet capacity is the overall economics of wind power as a function of capacity, such an analysis has not been attempted here due to the complexity of many inputs for which there is little public information including government subsidies, the structure of the base load and the cost of ancillary equipment for the integration of the wind fleet into the grid, such as standby dispatchable generating equipment. Instead we have chosen to focus on wind fleet efficiency as one of several important factors in any consideration of the wind power economics.

When considering a new wind investment, it can be assumed that it will generate roughly 35% of its nameplate capacity for an off shore investment and roughly 25% for an onshore investment. As may be seen in Figure 12 this will no longer be the case once the wind fleet has increased in size to the point where excess wind is shed. The load factor is now reduced to the slope of the appropriate curve in Figure 12, and this is seen to decrease progressively as the wind fleet increases in size. For investment purposes the appropriate load factor is no longer the nameplate load factor but the slope of the appropriate $GW_e$ vs. $GW_c$ curve of Figure 12. Because this is a measure of the increase in $GW_e$ for an increase in $GW_c$, we shall call this the wind fleet's marginal load factor.

We could use Figure 12 directly to calculate the marginal load factor for a new investment, but a much more efficient way of doing this is to derive from this figure a set of curves which generalise the relationship between the marginal load factor and wind fleet capacity for different levels of base generation. These new relationships are shown in Figure 14.

**Figure 14. Marginal load factor as a function of capacity and different levels of base generation**

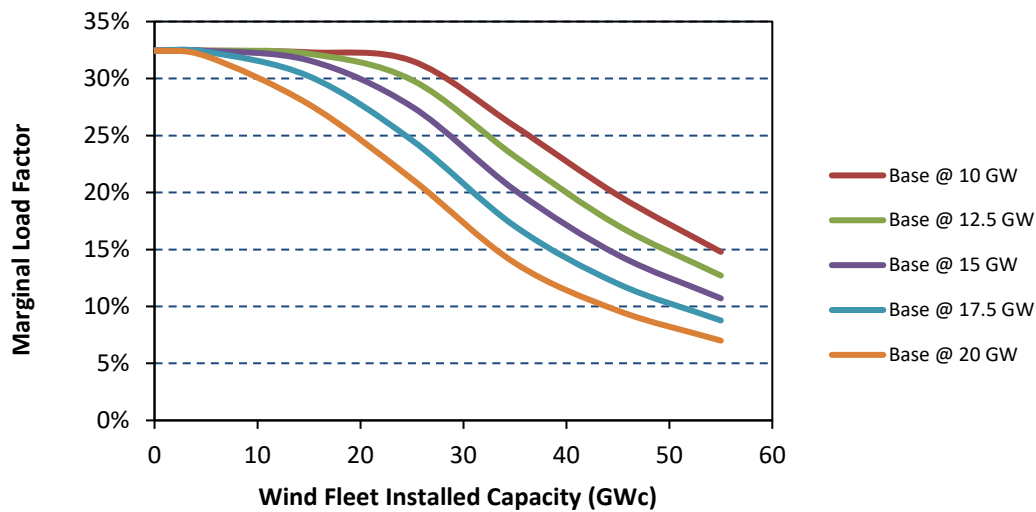

Rather than having to measure the gradients of the curves of Figure 12 for each case of interest, Figure 14 enables us to read off directly the marginal load factor for any wind fleet capacity and base generation. Thus for example, the wind fleet capacity at the end of 2015 was 13.6$GW_c$ and base generation about 12.5$GW_e$, so a new investment's marginal load factor in 2016 will be almost identical to its nameplate efficiency (or load factor) of 32.5%. Should the base generation remain the same but the wind fleet increase in size to 34.4$GW_c$, where the latter is the lowest anticipated wind fleet size as reported by the RAE, the marginal load factor will fall from 32.5% to 24%. Likewise, should the 2030 base generation and wind fleet capacity be 15$GW_e$ and 34.4$GW_c$ respectively, the marginal load factor will be 20%, equivalent to only 63% of its nameplate capacity; and if we assume that government reaches its aim of 16$GW_e$ of new nuclear generation by 2030 and that the base generation is 20$GW_e$,

Figure 14 indicates that a 34.4GW$_c$ wind fleet would have a marginal load factor of only 15%, less than half the nameplate load factor for the new investment.

Figure 14 confirms that off shore wind investment will become progressively more difficult to justify as the wind fleet increases in size. Even if there were no nuclear investment and the base generation falls to below 10GW$_e$, it is difficult to see how investment up to 50GW$_c$, the median from the report of the Royal Academy of Engineering (2014) for 2030, could be economically justified. Indeed the National Audit Office was highly critical of the agreed contractual arrangements for the 1.2GW$_e$ Hornsea off-shore wind farm (Gosden, 3 February 2016). Of particular concern to the office was its estimate that the project would require £4.2 billion in subsidy, an average of £240 million a year over the 15 year contract period. Consumers will be required to make up the difference between the current market price of £35/MWh and the guaranteed price of £140/MWh. Furthermore it must be remembered that this guaranteed price takes no account of the future cost of wind shedding, also to be borne by the consumer rather than the contractor.

### 6.2 Contribution of the Wind Fleet to Reducing Carbon Dioxide Emissions

During 2014 the UK wind fleet generated an average of 3.65GW$_e$ of electricity, thereby displacing displaced 3.65GW$_e$ of coal generation. Given that each GW$_e$ of coal generation is responsible for producing approximately 7.9 million tonnes per annum of carbon dioxide ($CO_2$) emissions, this is equivalent to a reduction in $CO_2$ emissions of 29 million tonnes. It is reported that electricity generation in 2014 was responsible for 122 million tonnes of $CO_2$, implying that in the absence of the wind fleet, the emissions would have been 151 million tonnes.

**Figure 15. Predicted CO2 emissions from electricity generation as a function of wind fleet capacity (base generation @ 15GW$_e$)**

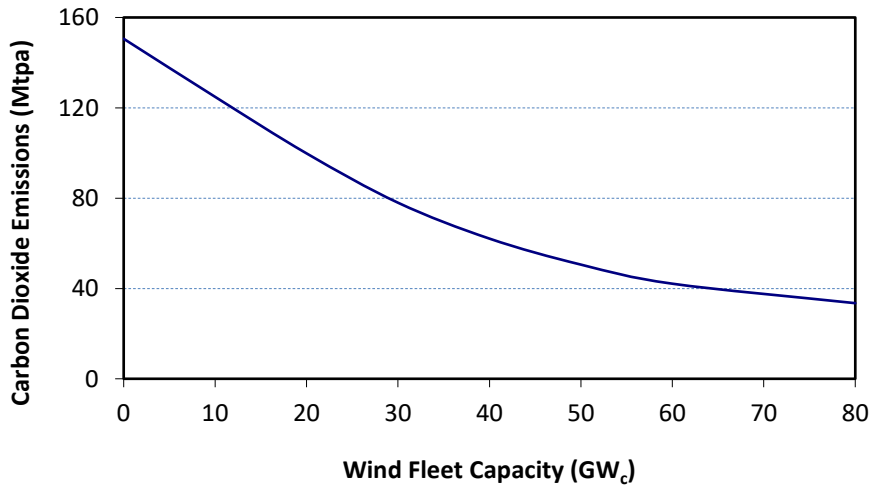

Using this value as the starting point and the expected wind fleet efficiencies as derived earlier, it is now possible to calculate the $CO_2$ emissions as a function of the wind fleet capacity. The predictions are shown in Figure 15. It is noted that the impact of the wind fleet in reducing emissions further falls rapidly above 35GW$_c$. Before 35GW$_c$ the impact is 2.57 million tonnes $CO_2$ per annum per GW$_c$; between 35GW$_c$ and 50GW$_c$, the effective efficiency is only 1.33 million tonnes $CO_2$ per annum per GW$_c$, and between 50GW$_c$ and 80GW$_c$ only 0.53 million tonnes $CO_2$ per annum per GW$_c$.

### 6.3 Security of Future Energy Supplies

In January 2016 there was a very public disagreement about the future security of UK energy supplies between the Confederation of British Industry and the Institution of Mechanical Engineers on the one hand, and the Department of Energy and Climate Change Secretary of State, Amber Rudd, on the other hand. In January 2016, The Times reported the publication of an open letter from the CBI (Pagnamenta, 26 January 2016). The article quoted the Chief Corporate Officer of Scottish Power, one of the letter's signatories, saying that:

> "Britain was facing an increasingly uncertain future in terms of energy supplies which have sapped confidence in the industry's investment climate".... "up to 25 sites for new gas-fired power stations had been granted planning permission in the UK, representing a potential investment of at least £16Bn …….. few if any of them are moving ahead because of lack of clear policy support and uncertainty over their commercial viability".

On the following day the Times reported that Amber Rudd had rejected the CBI's criticism, declaring that "we are clear that a range of energy sources such as nuclear, off-shore wind and shale gas all have roles to play in a low-carbon energy mix, powering our country and safeguarding our future economic security" (Pagnamenta, 27 January 2016).

A report from the Institution of Mechanical Engineers questioned whether it was now even possible to put in place the generating capacity needed in the next decade (Institution of Mechanical Engineers, 2016). The report noted that:

> "Under current policy, it is almost impossible for the UK electricity demand to be met in 2025"…" we neither have the time nor enough people with the right skills to build sufficient power plants. Electricity imports will put the UK's electricity supply at the mercy of the markets, weather, politics of other countries, making electricity less secure and less affordable".… "currently there are insufficient incentives for companies to invest in any sort of electricity infrastructure or innovation".

The data in Table 3 (UK's electricity generation by source) and Table 4 (nuclear closure dates) suggests that approximately 35GW$_e$ additional capacity might be needed by the late 2020s to compensate for the shutting down of coal stations and the loss of AGRs without nuclear replacements.

**Table 3. United Kingdom plant capacity (GW$_e$)**

| Year | 2010 | 2014 | Comment |
|---|---|---|---|
| Conventional coal | 35,315 | 24,838 | To be phased out by 2025 |
| CCGT | 34,026 | 33,784 | No published plan to increase CCGT capacity |
| Nuclear stations | 10,865 | 9,937 | Possibly down to only 1,198 after shut down of AGRs in the 2020s |
| Gas turbines and oil engines | 1,779 | 1,787 | |
| Hydro natural flow | 1,526 | 1,557 | Weather dependent |
| Hydro pumped storage | 2,744 | 2,744 | Max energy about 30GWh |
| Wind | 2,323 | 5,585 | Weather dependent |
| Other renewable | 1,896 | 4,747 | Biomass is the fastest growing other source, but may be limited by supplies of 'green' woodchip |
| **Total** | **90,473** | **84,987** | |

Source: The data was derived from Department of Energy & Climate Change (2015).

With gas currently available at low prices, it might be tempting to argue that the quickest and most cost effective means of both decarbonising the grid and improving the security of future energy supplies would be to make the building of new gas fired stations a higher priority rather than extending the wind fleet. Indeed, according to the National Audit Office, it is particularly difficult to justify further increases in off-shore wind capacity, with the tariff of £140 per MWh, guaranteed for the next 15 years, to be paid to the 1.2GW$_e$ Hornsea wind farm being just one example of the high cost for the technology (Gosden, 3 February 2016).

**Table 4. Currently anticipated nuclear reactor closure dates**

| Plant | Type | Capacity (GW$_e$) | First Power Date | Expected Closure Date |
|---|---|---|---|---|
| Dungeness B 1 & 2 | AGR | 520 & 520 | 1983, 1985 | 2028 |
| Hartlepool 1 & 2 | AGR | 595 & 585 | 1983, 1984 | 2024 |
| Heysham 1 1 & 2 | AGR | 580 & 575 | 1983, 1984 | 2024 |
| Heysham II 1 & 2 | AGR | 610 & 610 | 1988 | 2030 |
| Hinkley Point B 1 & 2 | AGR | 475 & 470 | 1976 | 2023 |
| Hunterston B 1 & 2 | AGR | 475 & 485 | 1976, 1977 | 2023 |
| Torness 1 & 2 | AGR | 590 & 595 | 1988, 1989 | 2030 |
| Sizewell B | PWR | 1,198 | 1995 | 2035 |
| **Total** | | **8,882** | | |

Source: World Nuclear Association (2015)

However, an argument against investing in gas generation to the exclusion of wind investment is that there is no guarantee that cheap gas will be in plentiful supply in future. We have already seen the Fukushima nuclear disaster in 2012 causing worldwide LNG shortages, and it is easy to suggest a number of scenarios under which future LNG supplies might be restricted, including political instability in one of the major LNG exporting countries, technical problems at one of the UK's import terminals or lack of gas storage (the UK has considerably less gas storage capacity than most of its continental neighbours). The UK has for many years adopted a policy of diversifying its energy sources, and a solution might be to both extend the wind fleet to the point at which its efficiency drops through wind shedding (around 25GW$_c$), but also building sufficient new gas generation capacity to keep the lights on in the late 2020s.

Although gas fired generation has lower capital costs than other forms of generation, the Economist pointed out that the government subsidies designed to stimulate the investment in renewables have undermined investment in gas (The Economist, 2015). Subsidised wind has been given favoured access to the grid but coal, the cheapest form of dispatchable generation, has been preferred to cleaner gas generation. What has undermined the investment in gas generation in recent years has been the under-utilisation of existing capacity, apparent when Table 3 and Table 4 are compared; gas generation utilisation fell from 58% in 2010 to 33% in 2014. According to the Economist, Germany has experienced a similar phenomenon, with wind farms and solar panels proliferating, but "dirty" coal and lignite generation increasing at the expense of "cleaner" gas generation (The Economist, 2015). Since government subsidies have been successful in stimulating investment in renewables, it ought to be possible for government to redress the current imbalance between coal and gas should it have the will to do so.

## 7. Conclusion

Although the shedding of excess wind power will become unavoidable as the wind fleet increases in size, there is little published information on the effect that this shedding will have on the overall wind fleet efficiency. This study addresses this issue through mathematical modelling.

An issue which has an important bearing on the upper economic limit of future UK wind fleets is whether generation which is surplus to requirement may be stored or transmitted for use in other countries. It is concluded that the size and short term duration of wind surpluses from large wind fleets makes it impractical to use of any technical means either currently available or likely to be available in the medium term to use these surpluses beneficially. Wind generation which is surplus to requirement will be shed, and it is the progressive shedding of more wind as the wind fleet increases in size which will put a practical upper limit on the size of the wind fleet.

To model the efficiency of future large UK wind fleets, wind generation records for the years 2013, 2014 and 2015 were downloaded from the web a week at a time. This data was then extrapolated to calculate how much wind would have been generated had the wind fleet capacity not been as in 2013, 2014 and 2015, but a range of capacities from $10GW_e$ to $80GW_c$. The model then calculated, for each 5-minute interval, how much of this wind could be used by the grid, the end result being three wind generation curves derived separately from 2013, 2014, and 2015 data. These were found to be sensibly the same, as shown in Figure 11.

An important and perhaps surprising prediction of the model is that the wind generation curve is almost insensitive to the size of the cyclic component of daily demand. The generation curve derived from 2014 data should therefore remain valid even if the future addition of smart equipment to the grid substantially reduced the size of the cyclic component of daily demand (Figure 13). Some compensation for changes in the proportions of on-shore to off-shore wind will be necessary but is easily effected, as described in Section 5.3.

The level of base generation turns out to be the most important determinant of the efficiency of the wind fleet, and it is argued that decrease in wind fleet efficiency will determine the upper economic limit of the wind fleet. Figure 14 shows how the wind fleet efficiency is expected to decrease progressively as the wind fleet increases in size, and confirms the suggestion of Lady Barbara Judge, chairwoman of the Institute of Directors that:

> "without cheap energy storage, at present a technical impossibility, intermittent renewables will be hard pressed to contribute much more than 25 per cent of our electricity".
> (Judge, 11 April 2016)